\documentclass[prc,twocolumn,superscriptaddress,showpacs,amssymb,amsmath,amsfonts,aps]{revtex4}
\usepackage{graphicx}
\usepackage{dcolumn}
\usepackage{epsfig}
\begin{document}
\title{Electroexcitation of nucleon resonances at $Q^2=0.65~(GeV/c)^2$\\
 from a combined analysis of single- and double-pion electroproduction data}

\newcommand*{\JLAB }{ Thomas Jefferson National Accelerator Facility, Newport
News, Virginia 23606, USA}
\affiliation{\JLAB }

\newcommand*{\YEREVAN }{ Yerevan Physics Institute, 375036 Yerevan, Armenia}
\affiliation{\YEREVAN }

\newcommand*{\MSU }{ Skobeltsyn Nuclear Physics Institute
at Moscow State University, 119899 Moscow, Vorobevy gory, Russia}
\affiliation{\MSU }

\author{I.G.~Aznauryan}
     \affiliation{\JLAB}\affiliation{\YEREVAN}
\author{V.D.~Burkert}
     \affiliation{\JLAB}
\author{G.V.~Fedotov}
     \affiliation{\MSU}
\author{B.S.~Ishkhanov}
     \affiliation{\MSU}
\author{V.I.~Mokeev}
     \affiliation{\JLAB}\affiliation{\MSU}

\noaffiliation

\thispagestyle{empty}

\begin{abstract}
{Data on single- and double-charged pion electroproduction off protons are 
successfully described in the second and third nucleon resonance regions
with common $N^*$ photocouplings. 
The analysis was 
carried out using separate isobar models for both reactions. 
From the combined analysis
of two exclusive channels,
the $\gamma^* p\rightarrow N^{*+}$ helicity amplitudes are obtained
for the resonances $P_{11}(1440)$, $D_{13}(1520)$, $S_{31}(1620)$, 
$S_{11}(1650)$, $F_{15}(1680)$, $D_{33}(1700)$, $D_{13}(1700)$,
and $P_{13}(1720)$ at $Q^2=0.65~(GeV/c)^2$. }
\end{abstract}

\maketitle

\section{Introduction}

Our knowledge of electromagnetic excitations of nucleon resonances
is mostly given by single-pion photo- and electroproduction experiments 
\cite{1}.
The only exception is the $S_{11}(1535)$ resonance, which is strongly
revealed in photo- and electroproduction of $\eta$ and is well investigated
in these processes (see, for example, Refs. \cite{2,3,4,5,6,7,8,9}).
Although with increasing masses, the couplings of the resonances
to multi-pion channels become increasingly significant, and for
many resonances dominant, the scarcity of experimental data
made it impossible to investigate properties of nucleon
resonances in the
photo-and electroproduction reactions with final states different
from $\pi N$ and $\eta N$.

The situation changed recently, when 
precise  measurements of single- and double-pion
electroproduction were carried out  
at Jefferson Lab
with the CEBAF Large Acceptance Spectrometer (CLAS).
In double-charged
pion electroproduction on protons,
the $\pi^{+}p$, $\pi^{-}p$, $\pi^{+}\pi^{-}$
invariant mass distributions, c.m.s. $\pi^{-}$
angular distributions, and total cross sections
were obtained  in the second, third,
and partly fourth resonance regions 
at $Q^2$ from $0.5$ to $1.5~(GeV/c)^2$ \cite{10,11}.  
At these $Q^2$, the CLAS measurements of single pion
electroproduction on protons include
differential cross sections 
and  polarized
beam asymmetries for $\pi^0$ and $\pi^+$ electroproduction \cite{11,12,13,14,15}.
Complemented with older DESY and NINA data
\cite{16,17,18,19,20},
 $\pi^0$ and $\pi^+$ cross sections extend from threshold
to the third resonance region.
Polarized
beam asymmetry measurements cover first and second resonance regions.

Using these measurements, we have carried out a combined analysis of
single- and double-pion electroproduction data at $Q^2=0.65~(GeV/c)^2$
in the second and third resonance regions. For this 
kinematics, the most complete set of data for the
two channels is available.
In this paper, we report on the results 
of this analysis, which 
was carried out using independent isobar models for each channel, but with common
$N^{*}$ photocouplings.

Reliable description of non-resonant mechanisms, as well as separation
of resonant and background contributions
represent fundamental problems in $N^{*}$ studies. Presently,
non-resonant processes can be treated only at a phenomenological level.
Reliability of background description and resonance/background 
separation can be put to the test in combined 
analyses of major exclusive channels.
Single- and double-pion channels 
in meson photo- and electroproduction
account  for almost 90\% of the total  
cross section in the $N^{*}$ excitation region. 
In addition, these production channels have completely 
different backgrounds. 
A successful description of these channels combined 
would confirm the reliability 
of the background description and
resonance/background separation. Therefore, this  
way we expect to obtain the most accurate values
for $N^{*}$ photocouplings with considerably reduced 
model uncertainties arising from
the phenomenological separation of resonances and background.

The analysis of two exclusive channels of
pion electroproduction on protons
has been performed within isobar models which are presented
in Section 2.
The analysis and the results are presented
in Section 3. In this section, we will demonstrate
that  the two main exclusive channels 
can be described with the same values
of $N^{*}$ photocouplings for the resonances
from the second and third resonance regions. The
sets of the $\gamma^*p\rightarrow N^{*+}$ helicity amplitudes
will be found that
give minimal $\chi^{2}$ values, fitting 
single-
and double-pion electroproduction data simultaneously.
The conclusions are presented in Section 4.

\section{The analysis tools}
\subsection{Single pion electroproduction}
The analysis of single pion electroproduction data has been
performed within the isobar model 
presented in Refs. \cite{8,21}.
In Ref.  \cite{21}, this model was successfully used
to describe multipole amplitudes of the GWU(VPI)
\cite{22,23} partial-wave analysis of $\pi$ photoproduction data.
It was also successfully used for the analysis
of CLAS data on $\pi$ electroproduction
at $Q^2=0.4$ and $0.65~(GeV/c)^2$ \cite{8}.
The model consists of resonance contributions parametrized in
the Breit-Wigner form and non-resonance
background built
from the Born term (the $s$ and $u$ channel nucleon exchanges
and $t$ channel $\pi$ contribution) and the $t$ channel
$\rho$ and $\omega$ contributions. The background is unitarized
in the $K$-matrix approximation. With increasing energy,
the background transforms into the amplitudes in the Regge
pole regime. In the present analysis the background was fixed in the way
described in Ref. \cite{8}, where at  $Q^2=0.65~(GeV/c)^2$
the analysis of the same data was done, focusing on the
$N^*$ mass region below $1.54~GeV$.

\subsection{Double-pion electroproduction}
The  analysis of  two-pion electroproduction data was
performed within the 
JLAB-MSU (Jefferson Lab -Moscow State University) isobar model \cite{24}. 
It incorporates significant 
improvements over the approach of Refs.
\cite{25,26}, which was used in the analysis of CLAS $2\pi$
electroproduction data in Ref.
\cite{10}.

In the initial version of the isobar model \cite{25,26}, 
double-charged-pion 
production was described by 
superposition of the following quasi-two-body channels with formation and
subsequent decay of unstable particles in the intermediate states:
\begin{eqnarray}
         \gamma p \rightarrow \pi^{-} \Delta^{++} \rightarrow
\pi^{-}  \pi^{+} p, \\
     \gamma p \rightarrow \pi^{+} \Delta^{0} \rightarrow
\pi^{+} \pi^{-} p,  \\
     \gamma p \rightarrow \rho^{0} p \rightarrow
\pi^{+} \pi^{-} p. 
\end{eqnarray}
Remaining (residual) mechanisms  were 
parametrized as 3-body phase 
space with the amplitude fitted to the data. 
This amplitude was a function of photon 
virtuality $Q^2$ and
the invariant mass of final hadronic system W only. In this 
approach, we were able to reproduce
the main features of integrated cross sections, as well as
invariant masses
$M_{\pi^+\pi^-}$, $M_{\pi^+ p}$
of $\pi^{+}\pi^{-}$, $\pi^{+}p$ systems, and  $\pi^{-}$ 
angular distributions.

The data on $\pi^{-}p$ invariant mass
distributions revealed evidence for the new isobar channel \cite{24}:
\begin{eqnarray}
     \gamma p \rightarrow  \pi^{+} D^0_{13}(1520) \rightarrow
\pi^{+} \pi^{-} p.
\end{eqnarray}
This mechanism allows us to describe an excess 
of strength measured in the $\pi^{-} p$ invariant 
mass distributions 
with respect to those 
calculated in the initial version 
of the JLAB-MSU model. This extra strength in the data, located around $1.52~GeV$ in 
$\pi^{-} p$ invariant mass
distributions, was clearly seen for all measured W bins kinematically 
allowed for the reaction (4).
The production amplitudes for the first three quasi-two-body 
mechanisms (1-3) were treated as sums
of $N^{*}$ excitations in the $s$-channel and non-resonant mechanisms
described in Refs. \cite{25,26}.
The quasi-two-body  mechanism (4) was
entirely treated as a non-resonant process \cite{24}.
In reactions (1-3), all well established 4-star resonances
with observed decays to the two-pion
final states were included as well as the 3-star states
$D_{13}(1700)$, $P_{11}(1710)$, $P_{33}(1600)$. For the latter state
a 1.67 GeV mass was obtained in our fit. This value
is in agreement with the results of recent analyses of the
$\pi N$ scattering experiments \cite{27}.
The electromagnetic couplings for the
$\gamma p \rightarrow N^{*+}$ vertices were
fitted to the data.
Hadronic coupling constants for $N^*\rightarrow \pi \Delta$
and $\rho p$ decays were taken from the analyses
of experiments with hadronic probes in the way described in Ref. \cite{25}.
The hadronic parameters for
a possible new baryon state $3/2^{+}(1720)$,
which has manifested itself
in the CLAS $2\pi$ photo- and electroproduction data
\cite{10,28}, were determined from a fit to the electroproduction data.
The hadronic couplings for $P_{13}(1720)$,
$D_{13}(1700)$, $P_{33}(1600)$
were also found from  fits to the CLAS $2\pi$ electroproduction data.

\begin{figure*}[tb]
\begin{center}
\epsfig{file=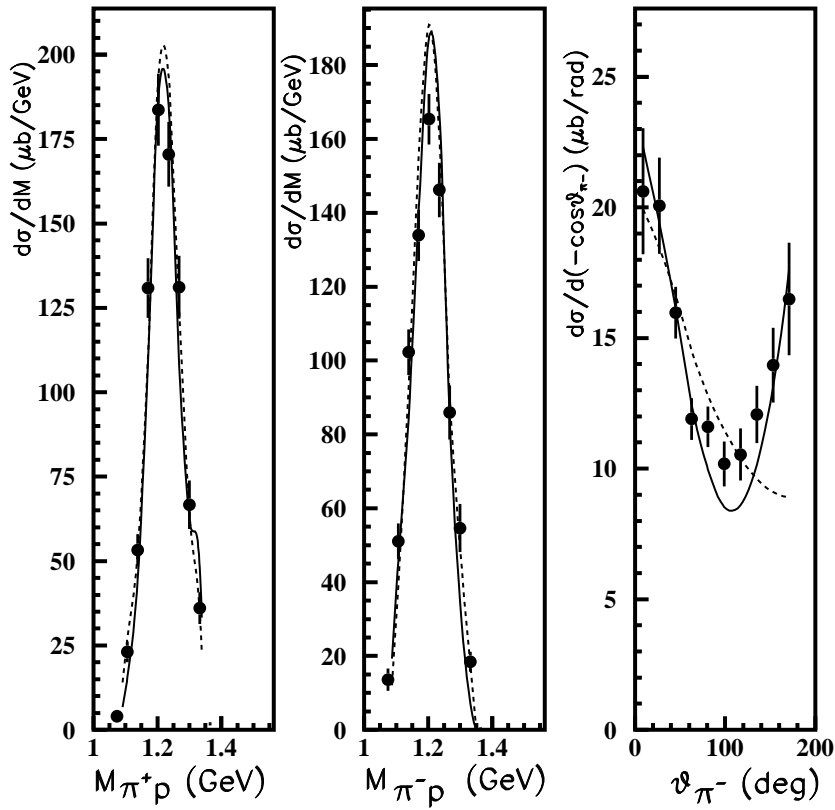,width=12cm}
\end{center}
\caption{CLAS 2$\pi$ electroproduction data \cite{10} 
in comparison with the results of the fits within the
initial  (dashed lines) and improved (solid lines) JLAB-MSU
model. W=1.49 GeV, $Q^{2}$=0.65 $(GeV/c)^{2}$.}
\label{fig:fig1}
\end{figure*}
\begin{figure}[t]
\includegraphics[scale=0.45]{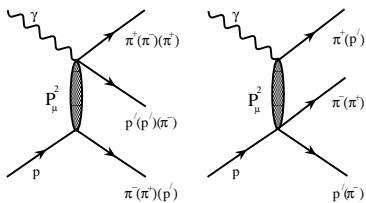}
\caption{Exchange background mechanisms for 2$\pi$ production, 
implemented in
improved JLAB-MSU model.}
\label{fig:fig2}
\end{figure}

In the improved JLAB-MSU model, we have also modified
the contribution from residual mechanisms. While the initial version of the 
model succeeded in reproducing
$\pi^{+}\pi^{-}$, $\pi^{+}p$ mass
distributions, it had shortcomings in the description of the c.m. 
$\pi^-$ angular distributions at low W (dashed lines in Fig. 1).
The 3-body phase space description of residual mechanisms was replaced 
by the set of exchange
terms shown in Fig. 2. This allowed a better 
description
of the $\pi^-$ angular distributions (solid lines in Fig. 1)
in the entire $Q^{2}$ range
covered by CLAS measurements.

The amplitudes of mechanisms shown in Fig. 2  
were parametrized in the Lorentz invariant form as follows:

\begin{equation}
A(W,Q^{2})\left[\varepsilon^{\gamma}_{\mu}\bar{U}_{p'}\gamma^{\mu}U_{p}
\right](P_{1}P_{2})e^{b\left( P^{2}_{\mu}-P^{2}_{\mu,min }\right)},
\end{equation}
where  $\varepsilon^{\gamma}_{\mu}$   is the photon
vector, $U_{p},\bar{U}_{p'}$ are the spinors for the initial and
final protons, $ P^{2}_{\mu}$ is the square of the 4-momentum transfer in
the exchange process, and $P^{2}_{\mu,min}$    
is the minimal allowed value of this quantity.
In Eq. (5), we have introduced the scalar product of four momenta 
 $(P_{1}P_{2})$ for the
pairs of  particles produced  in the diagram vertices;
this allowed us to reproduce
the shapes of the corresponding invariant mass distributions. Following
the diffractive ansatz, successfully used for the description of 
vector meson photo- and electroproduction, we have parametrized
the propagators in the exchange diagrams shown in Fig. 2 as 
exponential functions 
of $ P^{2}_{\mu}-P^{2}_{\mu,min }$.
Parameter b is the constant value found from the
fit to the data. The overall strength for each exchange mechanism
was adjusted to the data. We required smooth, structureless behavior of
cross-sections from these mechanisms as a function of W. 
 The sub-processes with $\pi^{-}$ in lower vertices of the Fig. 2 diagrams 
create backward peaks in the $\pi^{-}$ angular
distributions. These are mechanisms which allowed us to reproduce angular
distributions at backward angles in the entire kinematics region
covered by the CLAS measurements. The
sub-processes with $\pi^{-}$ in the upper vertices were necessary to provide 
good description
of mass distributions, as well as of the $\pi^{-}$
angular distributions at forward angles at $W\ge 1.7 GeV$.

With the improved JLAB-MSU model,  
a much better description of the data was obtained.

\section{The analysis and results}
The database for single-pion electroproduction includes
both recent CLAS and older
NINA and DESY data on protons, in particular:

(a) the CLAS data on $\pi^0$  cross sections
($W=1.1-1.52~GeV$, $E_e=1.645~GeV$ and
$W=1.1-1.68~GeV$, $E_e=2.445~GeV$) \cite{12},
 $\pi^+$
cross sections ($W=1.1-1.41~GeV$) \cite{13},
and polarized beam asymmetry
in $\pi^0$
and $\pi^+$ electroproduction ($W=1.1-1.58~GeV$) \cite{14,15};

(b) the DESY and NINA data on $\pi^0$
and $\pi^+$ differential cross sections
at $W=1.4-1.78~GeV$ \cite{16,17,18,19,20}.

Therefore, the data used in the analysis of single-pion
electroproduction include first,
second, and third resonance
regions. The data in the third resonance region
consist of CLAS data on  $\pi^0$  differential cross sections,
and the DESY and NINA data which, unlike CLAS measurements with
full angular coverage, extend mostly
over limited ranges of angles. 9870 data points were analyzed in 
the one-pion channel.

The two-pion electroproduction  data
are all from CLAS. They cover the W-range
from 1.4 to 1.9 GeV, and are composed of 20 W-bins with 0.025 GeV bin width.
Each bin contains data on $\pi^{+}\pi^{-}$, $\pi^{+} p$, $\pi^{-} p$ invariant
mass (8 points for each cross-section) and $\pi^{-}$ angular distributions
(10 points). Overall 680 data points were fitted.

In the analysis we have taken into account all 4 and 3 star resonances
from the second and third resonance regions.
The parameters of these resonances are listed in Table 1.
Along with the ranges taken from RPP \cite{29},
we have presented the values of masses and widths
used in the fits; these values were fixed.
We have also presented the values of branching ratios used
for extraction
of helicity amplitudes for the $\gamma^*N\rightarrow N^*$
transitions.
At $Q^2=0$, the couplings of the resonances $P_{33}(1600)$,
$D_{15}(1675)$ and $P_{11}(1710)$ to $\gamma N$ are small.
Our analysis  showed that
these resonances have minor contributions to the resonant  
electroproduction cross sections.
By this reason, the states $P_{33}(1600)$,
$D_{15}(1675)$ and $P_{11}(1710)$ are not listed in Table 1.
The photocouplings of the $P_{11}(1710)$ were taken 
equal to 0, since any
significant contribution from this state spoiled 
the shapes of $\pi^{+} \pi^{-}$ mass
distributions in 2$\pi$ electroproduction data.
The absolute values of the  helicity  
amplitudes for the $\gamma^*p\rightarrow P^+_{33}(1600)$ and
$\gamma^*N\rightarrow D^+_{15}(1675)$ transitions obtained in our analysis 
were below $0.02~GeV^{-1/2}$.
The $S_{11}(1535)$ resonance has small branching ratio
to the $\pi\pi N$ channel; it was included only in the
analysis of single-pion electroproduction
data. In that analysis, we have also included the 
$P_{33}(1232)$ resonance; its parameters were fixed according to the results
obtained in Ref. \cite{8}.

\begin{figure*}[tb]
\begin{center}
\epsfig{file=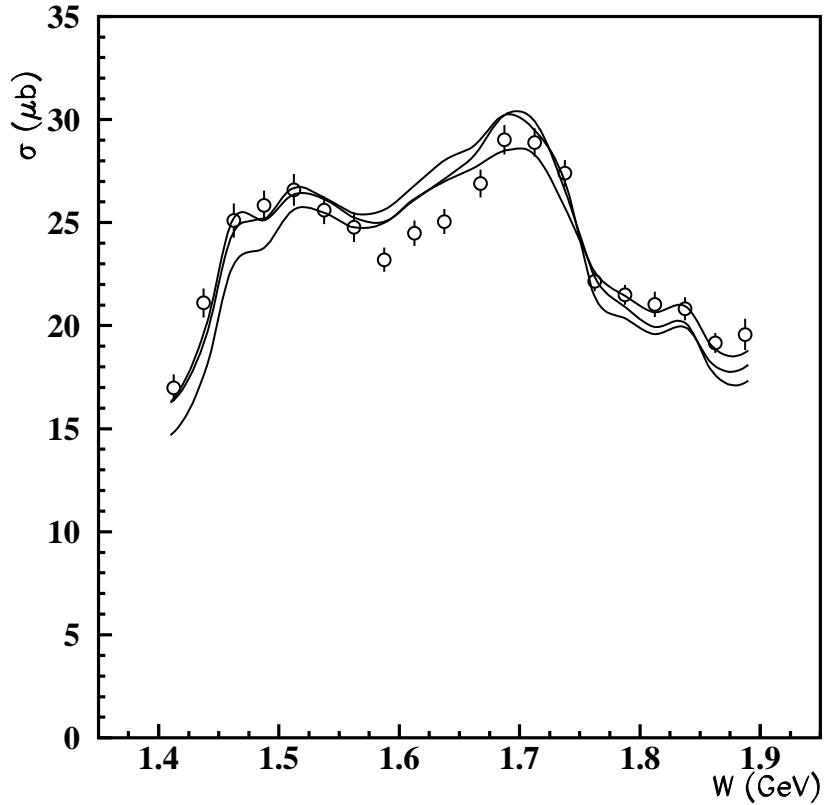,width=12cm}
\end{center}
\caption{\small{Description of CLAS data \cite{10} 
on total double charged pion
electroproduction cross sections with $N^{*}$ photocouplings 
taken from step 1.
Solid lines corresponds to the closest to the data calculated 
cross sections.}}
\label{fig:fig3}
\end{figure*}

The combined analysis of $\pi$ and 2$\pi$ electroproduction data
was made in several steps which are presented below.

\begin{figure*}[tb]
\begin{center}
\epsfig{file=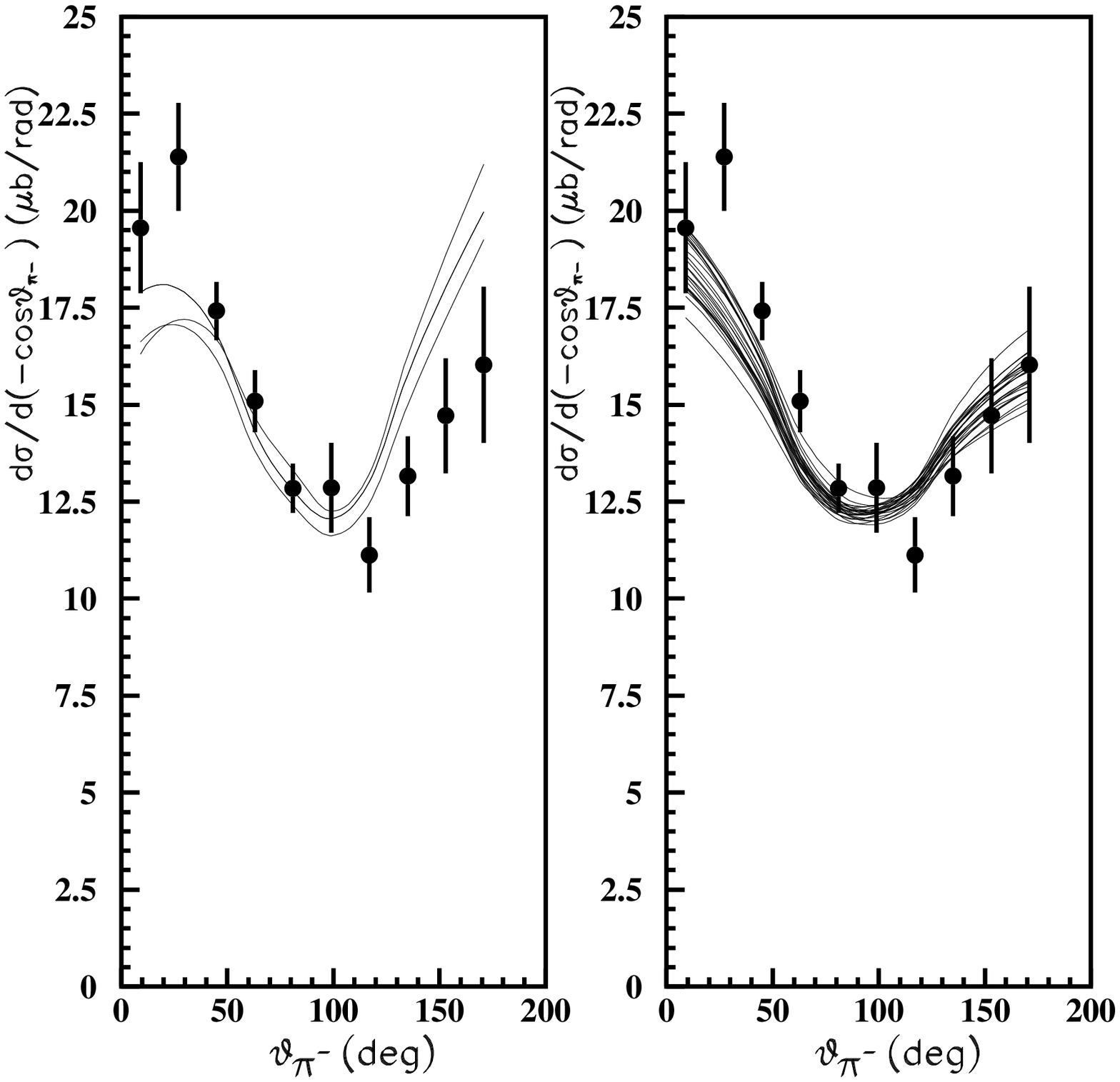,width=12cm}
\end{center}
\caption{\small{Description of CLAS 2$\pi$ electroproduction data \cite{10} 
on $\pi^-$ angular
distributions at W=1.71 GeV, $Q^{2}$=0.65 $(GeV/c)^{2}$. Left: closest to the data
cross sections calculated with $N^{*}$ photocouplings obtained in "1$\pi$
analysis". Right: cross sections with $N^{*}$ photocouplings found in "2$\pi$ analysis".}}
\label{fig:fig4}
\end{figure*}

\subsection{Step 1}
We started with the description of
single-pion electroproduction data. The observables 
mentioned above
were evaluated within the framework 
of the unitary isobar model \cite{8,21}.
The helicity amplitudes for the $\gamma^*p\rightarrow N^{*+}$
transitions were the only 
free parameters fitted to the data. 
The parameters of the model related to the background
(Born term and $\rho$ and $\omega$ contributions) were fixed
according to Ref. \cite{8}.
Masses, widths and $\pi N$ branching ratios of the resonances
were fixed at
the values given in Table 1.
Values of photocouplings and their uncertainties were obtained 
via minimization of 
$\chi^2$
over all data points using the MINUIT package  \cite{30}.  
The obtained results are presented in Table 2 in the columns
corresponding to "$1\pi~analysis$".
The value of $\chi^{2}/$(data points) over all
data was 1.19. The quoted errors are the fit uncertainties corresponding to the 
global $\chi^2$ minimum. 

The results for the  resonances $P_{11}(1440)$,
$D_{13}(1520)$ listed in Table 2, and for the $S_{11}(1535)$,
coincide with those obtained in Ref. \cite{8}, where at  $Q^2=0.65~(GeV/c)^2$
the analysis of the same data was done.

\begin{figure}[t]
\includegraphics[scale=0.45]{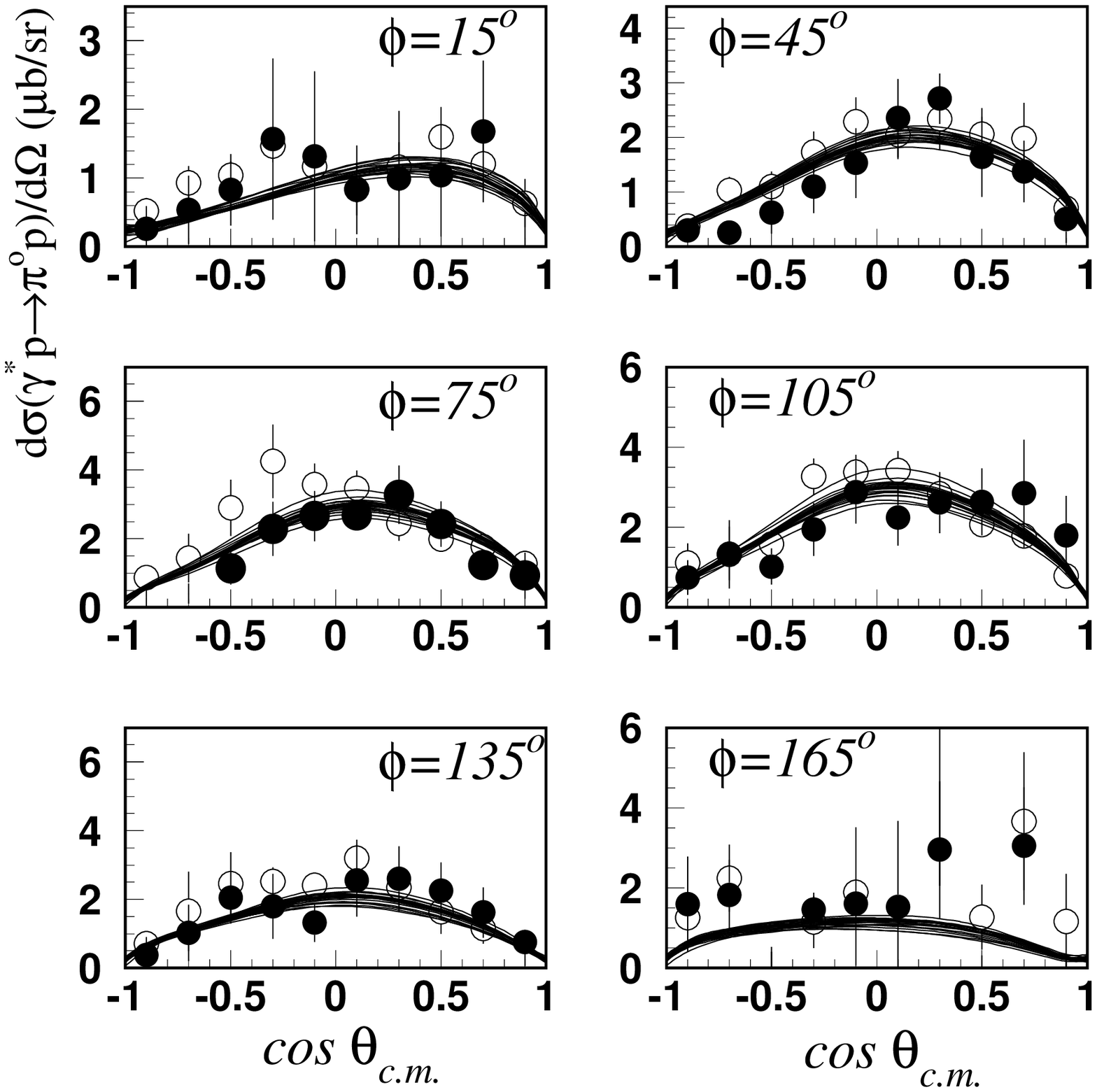}
\caption{ Differential cross section
for $\gamma^* p\rightarrow \pi^0 p$
at  $W=1.52~GeV$.
The data are from CLAS \cite{12}.
Open and solid circles correspond to measurements
with $E_e=1.645$ and $2.445~GeV$, respectively.
The  curves correspond to the sets of the
 $\gamma^* p\rightarrow N^{*+}$ helicity amplitudes
obtained in the final step of our analysis (step 3).}
\label{fig:fig5}
\end{figure}

\begin{figure}[t]
\includegraphics[scale=0.45]{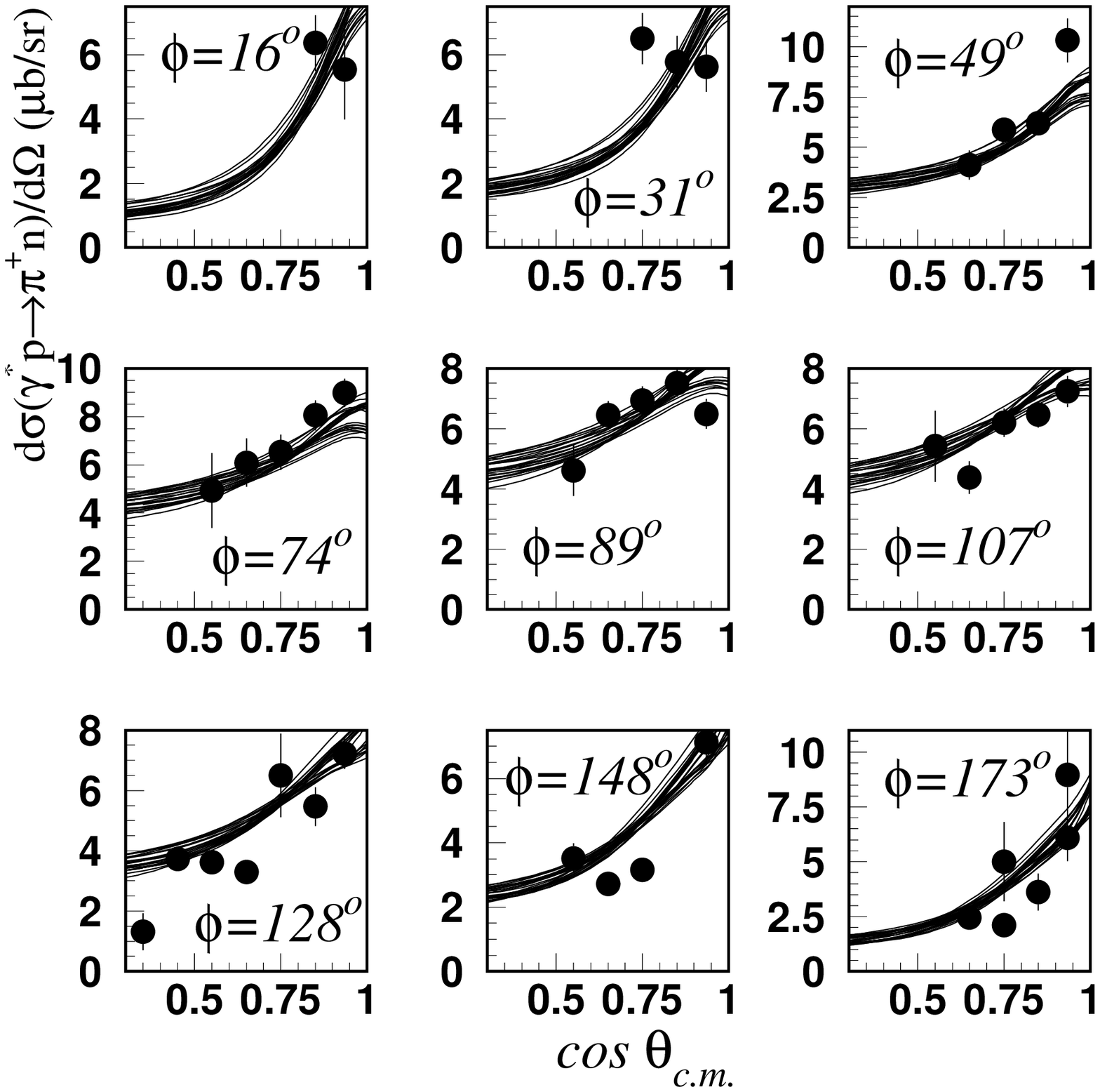}
\caption{ Differential cross section
for $\gamma^* p\rightarrow \pi^+ n$
at  $W=1.535~GeV$.
The data are from DESY \cite{18}.
The  curves correspond to the sets of the
 $\gamma^* p\rightarrow N^{*+}$ helicity amplitudes
obtained in the final step of our analysis (step 3).}
\label{fig:fig6}
\end{figure}

\subsection{Step 2}
Further, taking into account the results of the "$1\pi~analysis$",
the  2$\pi$ electroproduction data \cite{10} were analyzed
within the framework of the
improved JLAB-MSU model. The transverse electroexcitation
helicity amplitudes ${}_pA_{1/2}$, ${}_pA_{3/2}$
for the resonances with
$M<1.7~GeV$ were varied
around the values, obtained in step 1, 
because these resonances have large branching ratios to
the $\pi N$ channel, and therefore, their amplitudes
found in step 1 are good starting values for the $2\pi$ electroproduction data fit.
The only exception is 
the amplitude ${}_pA_{1/2}$ for the $P_{11}(1440)$,
as its values found in Ref. \cite{8}
using different approaches (isobar model
and dispersion relations) were significantly different.
In our analysis of 2$\pi$ electroproduction data, this amplitude 
was varied inside the range which overlapped the values reported in
Ref. \cite{8}.
For the resonances with $M\geq 1.7~GeV$, which decay mostly
to the $\pi\pi N$ channel, initial
values of transverse amplitudes were taken from the previous
analysis \cite{31} of 2$\pi$ electroproduction data 
within initial version of the JLAB-MSU model.
The initial values of all longitudinal amplitudes ${}_pS_{1/2}$
were taken from step 1. As the values
of ${}_pS_{1/2}$ for the resonances $D_{13}(1700)$ and $P_{13}(1720)$
found in step 1 were compatible with 0, 
they were taken equal to zero and kept unchanged
in the fitting procedure.

Among the resonances listened in Table 1, 
there are two strongly excited states in single-pion electroproduction 
which play an important role in the description of the data:
$D_{13}(1520)$ and $F_{15}(1685)$.
The resonance $D_{13}(1520)$ 
contributes almost half of the resonant cross sections
in the second resonance region 
and is responsible
for correct description of angular dependencies
of the cross sections and polarized beam asymmetries
in this region. 
$F_{15}(1685)$
gives large contribution to the resonant $\pi$ 
electroproduction cross section
in the third resonance region; its contribution is very important
for the description of angular dependencies
in this region.
These resonances were also  observed in 
$\pi\pi N$ channel with considerable excitation strengths \cite{31}. 
To study the sensitivity of photocouplings for 
the $D_{13}(1520)$ and $F_{15}(1685)$ to
single- and double-pion production data, we performed two separate fits of the 2$\pi$ 
electroproduction data
with different ranges of sampling for $N^{*}$ photocouplings.

Initial values of helicity amplitudes
were sampled according to normal
distribution.  
The variations $\sigma$  were taken equal to 30\% for all states, except 
$D_{13}(1520)$, $F_{15}(1685)$, and the ${}_pA_{1/2}$ amplitude 
for the $P_{11}(1440)$,
which was varied inside the range  obtained
in Ref. \cite{8}.
In the first fit (Fit A), the photocouplings of $D_{13}(1520)$ 
and $F_{15}(1685)$ were
varied inside uncertainties obtained at step 1. So 
in this fit, we assumed
that the $D_{13}(1520)$, $F_{15}(1685)$ photocouplings were driven by 
single-pion electroproduction. 
The second fit (Fit B)  was performed by
varying these photocouplings with
$\sigma$ equal to 30 \%.
The amplitudes of the exchange diagrams shown in Fig. 2 were varied in both
fits within 20\%, applying  W-independent multiplicative 
factors for each mechanism.
The range of calculated differential cross sections
overlapped entirely the measured cross sections, showing
that actual values of $N^{*}$
photocouplings
are inside intervals adopted for the photocoupling variations.
For each set of $N^{*}$ photocouplings,
the value of $\chi^{2}$  
was estimated.
Further, we have selected the sets of $N^{*}$ photocouplings,
corresponding to values of $\chi^{2}$,
when deviation between calculated
cross sections and the data were inside
the uncertainties of the measurements.
This minimization procedure for Fit A finally
left us with 16 sets of $N^{*}$ photocouplings with  
$2.81<~\chi^{2}/$(data point)~$<3$.
All these photocouplings allowed us to get
good description of $2\pi$ electroproduction  data.
A similar minimization 
procedure for Fit B
gave us additional 6 sets of $N^{*}$ photocouplings with
$2.83<~\chi^{2}/$(data point)~$< 2.95$.
So Fit B did not allow us
to improve the description for the 2$\pi$ electroproduction 
data set with respect to Fit A, 
where $D_{13}(1520)$ and
$F_{15}(1685)$ photocouplings were driven by the
data on single-pion electroproduction. 

The sets of $N^{*}$ photocouplings selected in Fit A and Fit B were averaged.
Mean values were assigned to the extracted $N^{*}$ photocouplings, while
dispersions were treated as the photocoupling uncertainties. The 
obtained amplitudes are presented
in Table 2 as "2$\pi$ analysis (step 2)" results. 
The ranges of photocouplings  were evaluated 
not from the global minimum for $\chi^{2}$,  
but by applying restriction
$\chi^{2}/$(data point)~$< 3$.
In this way we have achieved more realistic
accounting of the influence of data uncertainties on the 
extracted $N^{*}$ photocouplings. We also took into account
the possibility of having multiple compatible descriptions of
the data with close $\chi^{2}$. 

As seen from Table 2,
for the resonances with considerable decay rates to the $\pi$N final state 
($P_{11}(1440)$, $D_{13}(1520)$, $S_{11}(1650)$, $F_{15}(1680)$), 
as well as for $D_{33}(1700)$, which 
decays mostly to $\pi\pi N$, the
photocouplings extracted in the analyses of single- and double-pion 
electroproduction data are
compatible within errors and the uncertainty of 
the $P_{11}(1440)$ 
${}_pA_{1/2}$ amplitude extraction from $\pi$ electroproduction.

For the resonances $S_{31}(1620)$, $D_{13}(1700)$, and $P_{13}(1720)$
with considerable decay rates to the $\pi\pi N$ final state, the
photocouplings extracted in analyses of single- and double-pion 
electroproduction data are different.
To understand the importance of this difference for the description
of 2$\pi$ electroproduction data, we tried to fit these data using $N^{*}$  
photocouplings sampled within the ranges 
determined from single-pion data analysis. We selected
3 solutions with smallest $\chi^{2}$:
$\chi^{2}/$(data point)~$=$~3.4-3.9. These values of $\chi^{2}$ are significantly
larger than those obtained in unconstrained analysis of 
2$\pi$ electroproduction data.
The $2\pi$ electroproduction total cross sections corresponding 
to 3 selected solutions
are shown in comparison with the data 
in Fig. 3. 
It can be seen that 
there is an excess in the calculated
cross sections at W between 1.56 and 1.7 GeV,  which suggests that 
the values of the $S_{31}(1620)$ photocouplings derived from 
single-pion electroproduction are overestimated.
The reduction of the photoexcitation strength for this state 
to the value derived from the fit in the $\pi\pi N$  channel
is vital to reproduce  double pion electroproduction 
data at  $W=1.56-1.7~GeV$. 
As for the $D_{13}(1700)$ and $P_{13}(1720)$ photocouplings,
if we would restrict their values by the ranges obtained in $\pi$ 
electroproduction data analysis, we would fail to 
reproduce $\pi^-$ angular distributions
in the backward hemisphere as is shown in Fig. 4. On the left part of this 
plot we compare
the data with the cross sections calculated using
3 sets of $N^{*}$ photocouplings taken
from the results of the single-pion data analysis. 
Unrestricted fit of 2$\pi$ electroproduction data allows us to
reproduce very well the angular distributions at W 
around 1.7 GeV, as is shown in the right
part of Fig. 4. 

\begin{figure}[t], 
\includegraphics[scale=0.45]{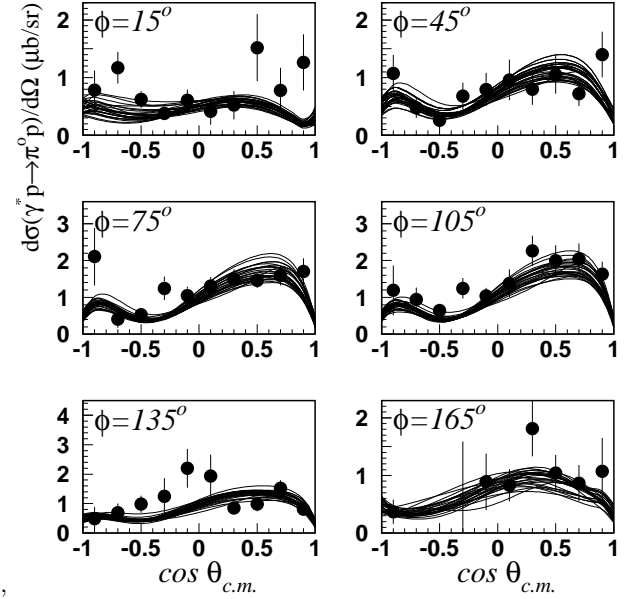}
\caption{The same as in Fig. 5 at W=1.68 GeV.} 
\label{fig:fig7}
\end{figure}

\begin{figure}[t]
\includegraphics[scale=0.45]{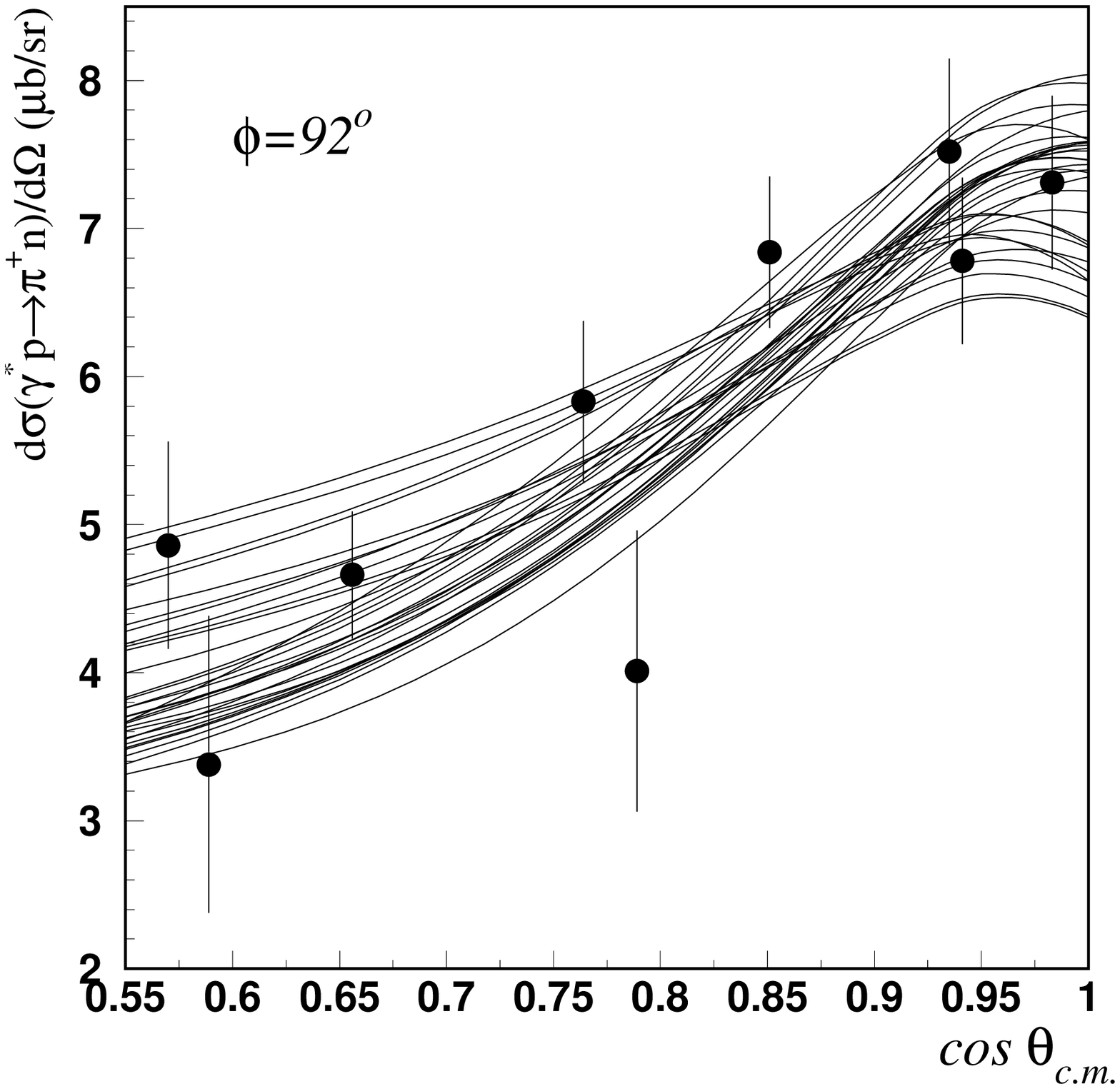}
\caption{ Differential cross section
for $\gamma^* p\rightarrow \pi^+ n$
at  $W=1.685~GeV$.
The data are from DESY \cite{20}.
The  curves corresponds to the sets of the
 $\gamma^* p\rightarrow N^{*+}$ helicity amplitudes
obtained in the final step of our analysis (step 3).}
\label{fig:fig8}
\end{figure}

\begin{figure}[t]
\includegraphics[scale=0.45]{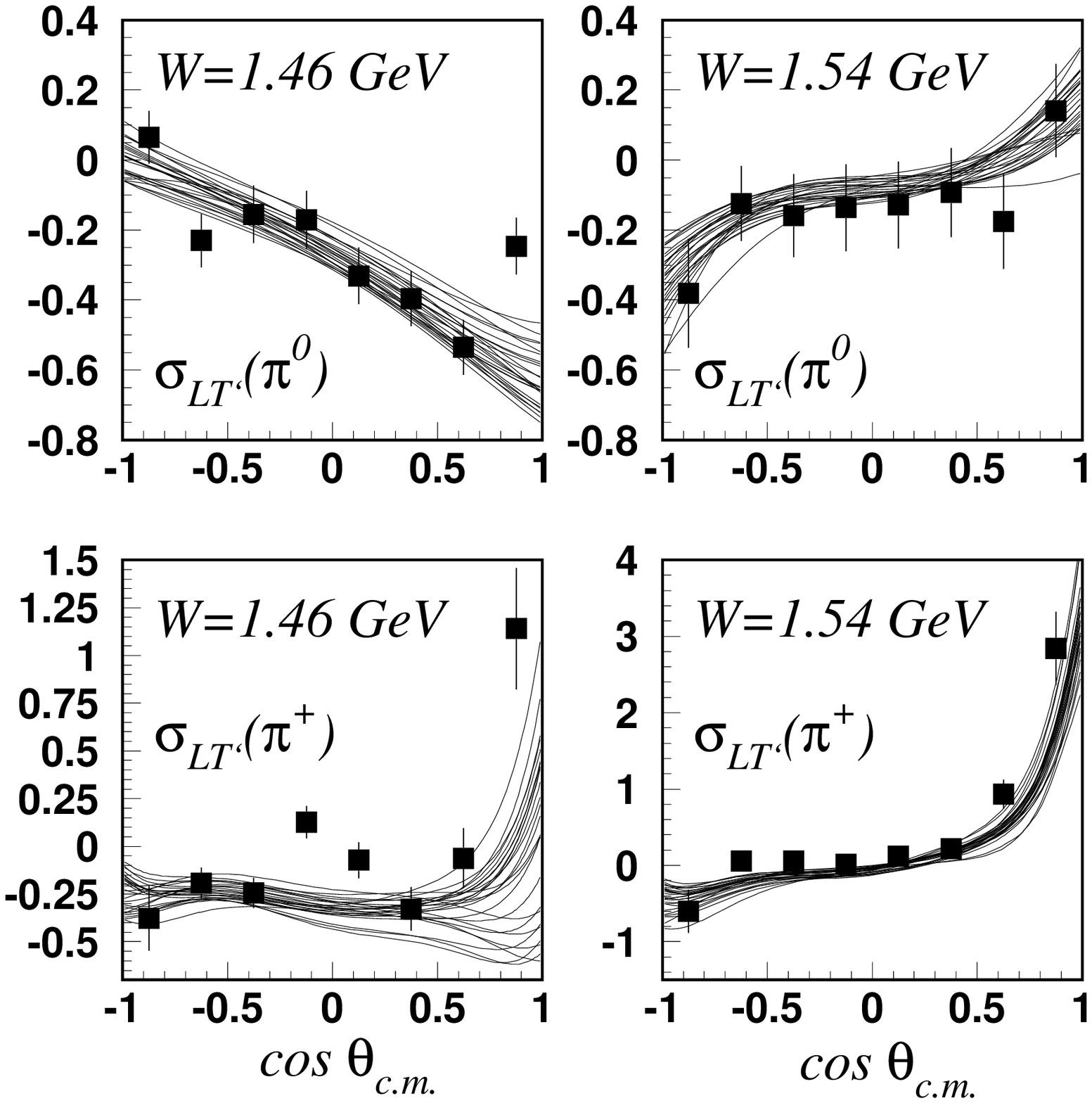}
\caption{ Structure function $\sigma_{LT'}$
for $\gamma^* p\rightarrow \pi^0 p$
and $\gamma^* p\rightarrow \pi^+ n$.
The data are from CLAS \cite{14,15}.
The  curves correspond to the sets of the
$\gamma^* p\rightarrow N^{*+}$ helicity amplitudes
selected in the final step of our analysis (step 3).}
\label{fig:fig9}
\end{figure}

\subsection{Step 3}
At this stage of the analysis we have finally found the ranges
of the $\gamma^*p\rightarrow N^{*+}$ helicity amplitudes which allow
us to obtain a good description
of electroproduction data in both $\pi N$ and $\pi\pi N$ channels.
First, for each of 16 sets of $N^{*}$ photocouplings
found at Fit A of the previous step, we have estimated the value of $\chi^{2}$
for single-pion electroproduction
from comparison between measured
and calculated observables.
It turned out that there are 3 sets of helicity amplitudes
that give the smallest values of $\chi^{2}$ simultaneously
in single- and double-pion electroproduction.
For these sets, we had $\chi^{2}/$(data point)$\simeq 1.24$
in the $\pi N$ channel, and $\chi^{2}/$(data point)$\simeq 2.85$
in the $\pi\pi N$ channel. 

The values of $\chi^{2}/$(data point) estimated for single-pion production
with 6 sets of $N^{*}$ photocouplings
found in Fit B were 
significantly larger than in  Fit A: $\chi^{2}/$(data point)$>1.31$, 
So only the sets
of $N^{*}$ photocouplings found in Fit A, where we restricted 
the variation of the $D_{13}(1520)$, $F_{15}(1680)$ photocouplings 
by the ranges determined
from $\pi$ electroproduction data, allow us to
get good description of both single- and
double-pion electroproduction observables. 

To determine the ranges of $\gamma^*N\rightarrow N^*$
helicity amplitudes
corresponding to the best description of
$\pi$ and 2$\pi$ electroproduction data combined, we have
repeated fitting both exclusive channels by applying photocoupling
variations around the mean values for
3 selected sets of $N^{*}$ photocouplings. 
Parameters $\sigma$ in normal distributions
were set 
at the dispersions equal to the ranges of
3 selected values of photocouplings
for each resonance except $S_{31}(1620)$,  
$D_{13}(1700)$, $D_{33}(1700)$, and $P_{13}(1720)$.  
As the photocouplings for these states
are strongly affected by the data from $\pi\pi N$ channel (Table 2),
their uncertainties in the variation procedure
were taken according to the results of "$2\pi$ analysis (step 2)".
W-independent
multiplicative factors for the Fig. 2 exchange mechanisms
in double-pion electroproduction
were varied within 5\% relative to the values obtained
at the previous step. The non-resonant
contributions for $\pi$ electroproduction
were fixed at the values used in step 1.
The above described procedure for combined fit
of $\pi N$ and $\pi\pi N$ channels was repeated.
Finally, we selected 30 sets of $N^{*}$ photocouplings which  provide
simultaneously good description of the data in both channels.
Average values and dispersions for selected sets of photocouplings
were attributed to the mean values
and uncertainties for $N^{*}$ photocouplings
that were extracted in the combined analysis.
They are presented in Table 2 in the columns
corresponding to "$1\pi-2\pi~analysis$".

\begin{figure}[t]
\includegraphics[scale=0.45]{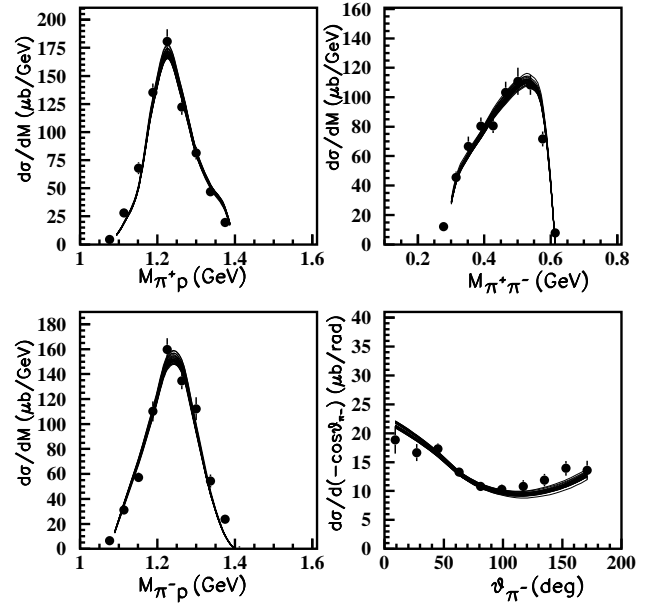}
\caption{$\pi^+ p$, $\pi^- p$, $\pi^+\pi^-$ invariant
mass and c.m.s. $\pi^-$ angular distributions in 2$\pi$ 
electroproduction at W=1.54 GeV. 
The data are from CLAS \cite{10}.
The  curves correspond to the sets of the
$\gamma^* p\rightarrow N^{*+}$ helicity amplitudes   
obtained in the final step of our analysis (step 3).}
\label{fig:fig10}
\end{figure}

\begin{figure}[t]
\includegraphics[scale=0.45]{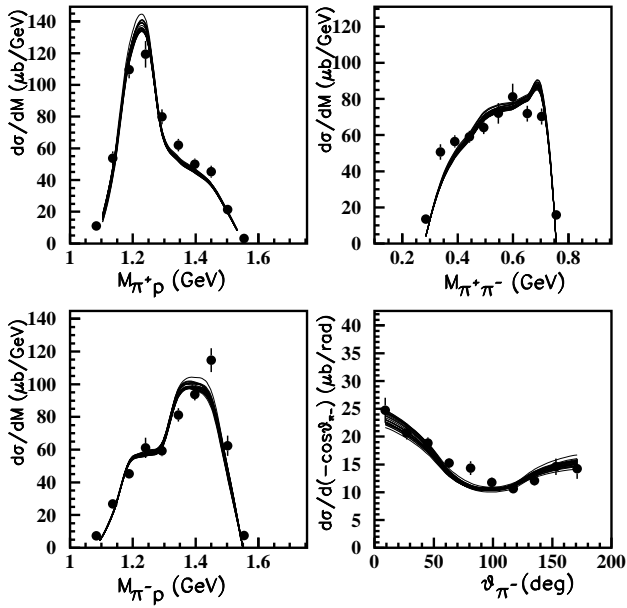}
\caption{The same as in Fig. 10 at W=1.69 GeV. }
\label{fig:fig11}
\end{figure}

\begin{figure}[t]
\includegraphics[scale=0.45]{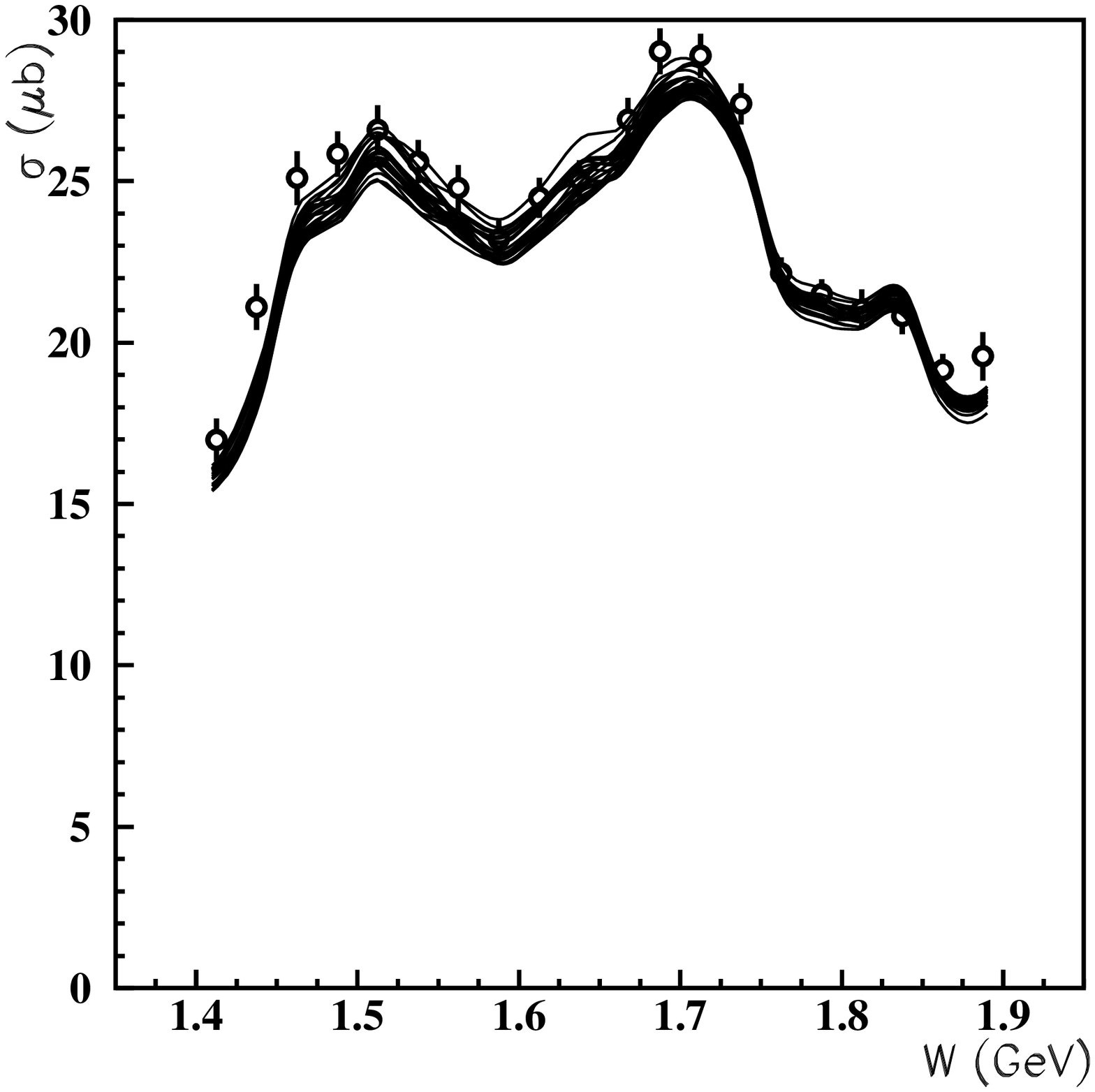}
\caption{Total cross section for 2$\pi$
electroproduction. The  curves corresponds to the sets of the
$\gamma^* p\rightarrow N^{*+}$ helicity amplitudes
obtained in the final step of our analysis (step 3). }
\label{fig:fig12}
\end{figure}

The quality of our description of single- and double-pion electroproduction data
is demonstrated in Figs. 5-12  where we present the results corresponding 
to the 
selected 30 sets of $N^{*}$ photocouplings
in comparison with experimental data. 
The results for  the invariant mass and angular distributions 
are presented in the centers
of the second and third resonance regions (Figs. 5-8,10,11).
The data for polarized
beam asymmetry in $\pi^0$ and $\pi^+$ electroproduction \cite{14,15}
extend from threshold to $W=1.58~GeV$, with bins equal to $0.04~GeV$.
In Fig. 9, we present the results for 
this observable at the energies,
which are
characteristic for the second resonance region;
the results are presented in the form of
the structure function $\sigma_{LT'}$. 
We have also presented the energy dependence of the 
2$\pi$ electroproduction total cross section (Fig. 12).  
From Figs. 5-12 it can be seen that the calculations made
with the 30 selected sets of $N^{*}$ photocouplings
are in good agreement with the data. The 
calculated curves are inside the data uncertainties, 
except for a few  data points.

As it follows from Table 2, the results obtained
in the final step of our analysis for the photocouplings of the resonances 
$D_{13}(1520)$, $S_{11}(1650)$, and
$F_{15}(1685)$
coincide within errors with those obtained 
separately in the analyses of single- and double-pion electroproduction data.
The errors of these photocouplings 
obtained in combined analysis 
are considerably lower
than those found in "2$\pi$ analysis", while comparable with "1$\pi$ analysis"
errors. Therefore, single-pion electroproduction data 
considerably affect the results on 
photocouplings
for low-lying states with masses below 1.7 GeV and sizable 
decay rates to the $\pi$N final state.
As for the ${}_pA_{1/2}$ $P_{11}(1440)$ amplitude,
$\pi$ electroproduction data have small sensitivity to its change
from the value obtained in the step 1 to one found in the final step.
As it follows from the results of the final step of our
analysis, the small error bar for
this amplitude, derived in step 1,
is related just to the way of evaluation of uncertainties
in $\pi$ electroproduction data fit, when only global minimum of $\chi^{2}$ was
taken into account. 

For the resonances with major decay rates to the 2$\pi$N channel 
($S_{31}(1620)$,
$D_{13}(1700)$, $D_{33}(1700)$, 
$P_{13}(1720)$), the photocouplings obtained in combined 
analysis 
coincide within the errors 
with the results of "2$\pi$ analysis".
However, the combined analysis allowed us
to considerably reduce the uncertainties
of these photocouplings compared with those derived in the analysis
of two-pion electroproduction data only.
The results of the combined analysis for  $S_{31}(1620)$,
$D_{13}(1700)$, $D_{33}(1700)$,
and $P_{13}(1720)$
are mostly different from those of "1$\pi$ analysis".
Nevertheless, as it can be seen from Figs. 5-9,
the photocouplings for these states obtained in combined 
analysis allow us to describe 
single-pion electroproduction data.

In Table 3, we compare our results for the transverse photocouplings
of the resonances from the $[70,1^-]$
multiplet with the predictions of single quark
transition model (SQTM) \cite{32}. This approach 
is based on the assumptions that
$SU(6)\bigotimes O(3)$ symmetry holds for the leading part
of confinement forces and 
only a single quark is affected in the 
electroexcitation of the $N^*$. For the states assigned
to the same $SU(6)\bigotimes O(3)$ multiplet,
this approximation allows us to
relate transverse $N^*$ photocouplings  
to limited number of parameters, which are three
in the case of the $[70,1^-]$ multiplet.  
In Ref. \cite{32}, these parameters were found from the experimental
data on the $S_{11}(1535)$ and $D_{13}(1520)$
photocouplings, and the predictions were made
for the transverse photocouplings of all other states 
assigned to $[70,1^-]$. 

As it can be seen, the photocouplings for all states
from $[70,1^-]$ obtained in our analysis, except $D_{13}(1700)$,
are in good agreement with SQTM predictions.  
SQTM results for $D_{13}(1700)$ were obtained using mixing
angle $\simeq 6^o$ between $|N^2,\frac{3}{2}^->$ and 
$|N^4,\frac{3}{2}^->$ configurations found in the analysis
of $D_{13}(1700)$ hadronic decays \cite{27}.
Accuracy of these data still remains poor. One can 
get
reasonable SQTM reproduction of our results for $D_{13}(1700)$,
if instead of $6^o$, a mixing angle $\simeq 20^o$ is used.
Good overall agreement of our results for the $N^*$ photocouplings 
with SQTM predictions strongly supports the complicated
dynamics of $N^*$ formation and
electroexcitation is determined mostly by
the $SU(6)\bigotimes O(3)$ spin-flavor-space symmetry,
and only 
a single quark is affected in the electroexcitation of the $N^*$. 

\section{Conclusion}

Recent CLAS and world data on  single- and double-charged pion
electroproduction off protons
are successfully described in the second and third resonance regions
using common values of $N^{*}$ photocouplings.
The analysis was carried out using isobar models
of Refs. \cite{8,21} and \cite{24,25,26} for  single- and double-pion
electroproduction, respectively.
The non-resonant mechanisms 
in these exclusive channels 
are completely different. Therefore, a
successful description of
$\pi N$ and $\pi\pi N$ channels combined
strongly suggests that: a) we achieved
reasonable treatment of non-resonant mechanisms;  
b) a phenomenological
separation between $N^{*}$ and background contributions 
made according to isobar models \cite{8,21} and \cite{24,25,26}
is reliable in both channels; and
(c) extracted $N^{*}$ photocouplings, which provide good  
description of all measured observables in these 
channels, 
have considerably reduced model uncertainties.

From our combined analysis,
the $\gamma^* p\rightarrow N^{*+}$ helicity amplitudes are extracted
for the resonances $P_{11}(1440)$, $D_{13}(1520)$, $S_{31}(1620)$,
$S_{11}(1650)$, $F_{15}(1680)$, $D_{33}(1700)$, $D_{13}(1700)$,
and $P_{13}(1720)$. 

Our results for the transverse photocouplings
of the resonances from the $[70,1^-]~SU(6)\bigotimes O(3)$
multiplet are in good agreement with the predictions of the single quark
transition model \cite{32}. Thus, we have strong evidence 
for $SU(6)$ as a leading symmetry of confinement forces
and single-quark transitions as dominant mechanism
in $N^*$ excitation by photons.

\widetext

\begin{table}
\begin{center}
\begin{tabular}{cccccccccccccccccc}
\hline
&&&&&&&&&&\\
$N^*$&&$M$&$\tilde{M}$&$\Gamma$&
$\tilde{\Gamma}$&$\beta_{\pi N}$&$\tilde{\beta}_{\pi N}$&
$\beta_{\pi\Delta+\rho N}$&$\tilde{\beta}_{\pi\Delta+\rho N}$&\\
&&&&&&&&&&\\
&&$(MeV)$&$(MeV)$&$(MeV)$&
$(MeV)$&$($\%$)$&$($\%$)$&
$($\%$)$&$($\%$)$&\\
&&&&&&&&&&\\
\hline
&&&&&&&&&&\\
$P_{11}(1440)$&&$1430-1470$&$1440$&$250-450$&
$350$&$60-70$&$60$&$20-30$&$25$&\\
&&&&&&&&&&\\
$D_{13}(1520)$&&$1515-1530$&$1520$&$110-135$&
$120$&$50-60$&$50$&$30-50$&$33$&\\
&&&&&&&&&&\\
$S_{31}(1620)$&&$1615-1675$&$1620$&$120-180$&
$150$&$20-30$&$25$&$70-80$&$75$&\\
&&&&&&&&&&\\
$S_{11}(1650)$&&$1640-1680$&$1650$&$145-190$&
$150$&$55-90$&$70$&$5-19$&$5$&\\
&&&&&&&&&&\\
$F_{15}(1680)$&&$1675-1690$&$1680$&$120-140$&
$130$&$60-70$&$65$&$8-30$&$17$&\\
&&&&&&&&&&\\
$D_{33}(1700)$&&$1670-1770$&$1700$&$200-400$&
$300$&$10-20$&$15$&$80-90$&$85$&\\
&&&&&&&&&&\\
$D_{13}(1700)$&&$1650-1750$&$1700$&$50-150$&
$100$&$5-15$&$10$&$85-95$&$90$&\\
&&&&&&&&&&\\
$P_{13}(1720)$&&$1650-1750$&$1720$&$100-200$&
$150$&$10-20$&$15$&$>70$&$85$&\\
&&&&&&&&&&\\
\hline
\end{tabular}
\caption{\label{tab1} List of masses, widths and branching
ratios of the investigated resonances.
The quoted ranges are taken from RPP \cite{29}.
The quantities labeled by tildes correspond to the values
used in our analysis.}
\end{center}
\end{table}

\begin{table}
\begin{center}
\begin{tabular}{ccccccccccccccccccccccc}
\hline
&&&&&&&&&&&&&&&&&&\\
$N^*$&&&&$1\pi$&&&&&&&&$2\pi$&&&&&&&&$1\pi-2\pi$&&\\
&&&&analysis&&&&&&&&analysis&&&&&&&&analysis&&\\
&&&&(step~1)&&&&&&&&(step~2)&&&&&&&&(step~3)&&\\
&&&&&&&&&&&&&&&&&&\\
&&&${}_pA_{1/2}$&${}_pA_{3/2}$&${}_pS_{1/2}$&
&&&&&${}_pA_{1/2}$&${}_pA_{3/2}$&${}_pS_{1/2}$&
&&&&&${}_pA_{1/2}$&${}_pA_{3/2}$&${}_pS_{1/2}$&\\
&&&&&&&&&&&&&&&&&&\\
\hline
&&&&&&&&&&&&&&&&&&\\
$P_{11}(1440)$&&&$4\pm 4$&$ $&$40\pm 4$&
&&&&&$21\pm 9$&$ $&$35\pm 6$&
&&&&&$21\pm 4$&$ $&$33\pm 6$&\\
$$&&&$(23\pm 4)$&$ $&$$&
&&&&&$$&$ $&$$&
&&&&&$$&$ $&$$&\\
&&&&&&&&&&&&&&&&&&\\
$D_{13}(1520)$&&&$-67\pm 3$&$62\pm 4$&$-38\pm 3$&
&&&&&$-62\pm 12$&$62\pm 11$&$-40\pm 7$&
&&&&&$-65\pm 4$&$62\pm 5$&$-35\pm 3$&\\
&&&&&&&&&&&&&&&&&&\\
$S_{31}(1620)$&&&$31\pm 3$&$ $&$-35\pm 3$&
&&&&&$13\pm 3$&$ $&$-25\pm 4$&
&&&&&$16\pm 4$&$ $&$-28\pm 3$&\\
&&&&&&&&&&&&&&&&&&\\
$S_{11}(1650)$&&&$43\pm 2$&$ $&$-12\pm 2$&
&&&&&$44\pm 16$&$ $&$-8\pm 2$&
&&&&&$43\pm 7$&$ $&$-6\pm 3$&\\
&&&&&&&&&&&&&&&&&&\\
$F_{15}(1680)$&&&$-38\pm 4$&$56\pm 2$&$-18\pm 2$&
&&&&&$-31\pm 8$&$52\pm 10$&$-14\pm 3$&
&&&&&$-32\pm 5$&$51\pm 4$&$-15\pm 3$&\\
&&&&&&&&&&&&&&&&&&\\
$D_{33}(1700)$&&&$58\pm 5$&$29\pm 3$&$-15\pm 5$&
&&&&&$49\pm 8$&$36\pm 8$&$-11\pm 4$&
&&&&&$44\pm 4$&$36\pm 4$&$-7\pm 4$&\\
&&&&&&&&&&&&&&&&&&\\
$D_{13}(1700)$&&&$-6\pm 8$&$-55\pm 7$&$0\pm 8$&
&&&&&$-19\pm 3$&$11\pm 2$&$0$&
&&&&&$-21\pm 2$&$10\pm 1$&$0$&\\
&&&&&&&&&&&&&&&&&&\\
$P_{13}(1720)$&&&$45\pm 6$&$-48\pm 7$&$0\pm 7 $&
&&&&&$56\pm 6$&$-62\pm 9$&$0$&
&&&&&$55\pm 3$&$-68\pm 4$&$0$&\\
&&&&&&&&&&&&&&&&&&\\
\hline
\end{tabular}
\caption{\label{tab2}  Helicity
amplitudes for the $\gamma^*p\rightarrow N^{*+}$ transitions
(in $10^{-3}GeV^{-1/2}$ units)
obtained at different steps of our analysis. Final results are presented
in the columns corresponding to "$1\pi-2\pi~analysis$".
In the brackets, we present the ${}_pA_{1/2}$ $P_{11}(1440)$ amplitude 
found in the dispersion relations analysis of pion
electroproduction in Ref. \cite {8}; 
this is the only result of this analysis
which is  different from that
found using isobar model. } 
\end{center}
\end{table}

\begin{table}
\begin{center}
\begin{tabular}{ccccccccccccccccccccccc}
\hline
&&&&&&&&&&&&&&&&&&\\
Resonance&&&&Our results&&&&&&SQTM\\
&&&&&&&&&&&&&&\\
&&&${}_pA_{1/2}$&${}_pA_{3/2}$&
&&&&&${}_pA_{1/2}$&${}_pA_{3/2}$&\\
&&&&&&&&&&&&&&&&&&\\
\hline
&&&&&&&&&&&&&&&&&&\\
$D_{13}(1520)$&&&$-65\pm 4$&$62\pm 5$&
&&&&&$-78\pm 10$&$63\pm 3$&\\
&&&&&&&&&&&&&&&&&&\\
$S_{31}(1620)$&&&$16\pm 4$&$ $&
&&&&&$18\pm 3$&$ $&\\
&&&&&&&&&&&&&&&&&&\\
$S_{11}(1650)$&&&$43\pm 7$&$ $&
&&&&&$43\pm 3$&$ $&\\
&&&&&&&&&&&&&&&&&&\\
$D_{33}(1700)$&&&$44\pm 4$&$36\pm 4$&
&&&&&$50\pm 5$&$35\pm 5$&\\
&&&&&&&&&&&&&&&&&&\\
$D_{13}(1700)$&&&$-21\pm 2$&$10\pm 1$&
&&&&&$-7\pm 2$&$6\pm 2$&\\
&&&&&&&&&&&&&&&&&&\\
\hline
\end{tabular}
\caption{\label{tab3}  
Our results for the resonances of the $[70,1^-]$ multiplet (in $10^{-3}GeV^{-1/2}$ units)
in comparison with the predictions of single quark transition model \cite{32}.}
\end{center}
\end{table}


\end{document}